\documentclass[aps,prb,twocolumn,groupedaddress]{revtex4-1}
\usepackage{graphicx}
\usepackage{dcolumn}
\usepackage{bm}
\usepackage{amsmath}
\usepackage{amssymb}
\usepackage{color}

\begin{document}


\title{Cup-to-vesicle transition of a fluid membrane with spontaneous curvature}

\author{Hiroshi Noguchi}
\email[]{noguchi@issp.u-tokyo.ac.jp}
\affiliation{
Institute for Solid State Physics, University of Tokyo,
 Kashiwa, Chiba 277-8581, Japan}

\date{\today}

\begin{abstract}
The disk-to-vesicle transition of a fluid membrane with no spontaneous curvature is well described by the competition
between edge line and curvature energies.
However, the transition of asymmetric membranes with spontaneous curvatures is not yet understood.
In this study, the shape of the fluid membrane patch with a constant spontaneous curvature
and its closing transition to a vesicle is investigated using theory and meshless membrane simulations.
It is confirmed that the (meta)stable and transient membranes are well approximated by spherical caps.
The membrane Gaussian modulus can be estimated from the cup shape of  membrane patches
as well as from the transition probability, although
the latter estimate gives slightly smaller negative values.
Furthermore, the self-assembly dynamics of membranes are presented, 
in which smaller vesicles are formed at higher spontaneous curvatures,
higher edge line tension, and lower density.
\end{abstract}

\maketitle

\section{Introduction}

Amphiphilic molecules self-assemble in aqueous solution to prevent contact between hydrophobic components
and water, forming mesoscale structures such as bilayer membranes and spherical and cylindrical micelles~\cite{safr94}.
Among these structures, the bilayer membranes consisting of lipid molecules are a basic structure of biomembranes;
therefore, they have received substantial research attention.
The morphologies of single-component lipid vesicles observed in experiments
can be theoretically understood by 
 minimization of the curvature energy with area and volume constraints
and can be reproduced by numerical simulations~\cite{seif97,hota99,yana08,svet14}.
However, biomembranes are more complicated than single-component membranes as
they consist of multiple types of  heterogeneously distributed lipids and proteins. 
Moreover, the  binding of proteins, such as BAR (Bin/Amphiphysin/Rvs) superfamily proteins,
can induce local membrane curvatures~\cite{mcma05,shib09,baum11,mcma11,suet14,joha15}.
As these proteins locally regulate the membrane shapes,
the effects of such membrane curvature  have been the focus of recent research~\cite{simu15a,nogu15b,nogu17,nogu17a,yu13,take17,phil09,lipo13,sari13,dasg17}.

This study clarifies the effects of spontaneous membrane curvature on vesicle formation.
A large bilayer membrane spontaneously closes into a vesicle;
this shape transition  occurs owing to competition between
the curvature energy and edge line energy.
For a single-component bilayer membrane,
the transition is well described theoretically by considering spherical-cap shapes 
as the transient geometry~\cite{from83,hu12}.
In this theory, the membrane is symmetric and has no spontaneous curvature, which
 is a reasonable assumption for lipid membranes
because the lipid can move between two layers by the diffusion through the membrane edges.
However, the protein binding can generate a finite spontaneous curvature.
Recently, Boye {\it et al.} reported that the binding of Annexin detaches supported lipid membranes
from the membrane edge and induces rolled membrane structures~\cite{boye17,boye18}.
Nevertheless, the effects of the spontaneous curvature for membranes with open edges have not been investigated so far.
Here, the shape of an isolated membrane patch 
and its transition into a vesicle are investigated using theory and simulation.
Protein binding can locally induce not only isotropic but anisotropic spontaneous curvatures, 
since the interaction can strongly depend on the domain axis~\cite{simu15a,nogu15b,nogu17,nogu17a,yu13,take17}.
The polymer anchoring and adhesion of spherical colloids 
can also induce isotropic spontaneous curvatures~\cite{phil09,lipo13,sari13,dasg17}.
For simplicity, this study considers that the membrane has a constant isotropic spontaneous curvature, $C_0$.

Section~\ref{sec:method} describes the simulation model and method.
Several types of simulation models have been developed to investigate membranes~\cite{muel06,vent06,nogu09}.
Here, a meshless membrane model is used where membrane particles self-assemble into a membrane.
One can efficiently simulate membrane deformation with topological changes 
in a wide range of membrane elastic parameters.
Section~\ref{sec:theory} describes the free energy of a spherical-cap-shaped membrane.
A previous theory~\cite{from83} for symmetric membranes ($C_0=0$)
is extended to consider spontaneous curvature.
Section~\ref{sec:close} describes the estimation method of the Gaussian modulus, $\bar{\kappa}$, 
using the closing probability proposed by Hu {\it et al.}~\cite{hu12}
This modulus is calculated  for the meshless membrane, revealing 
that it is applicable not only for $C_0=0$ but also for $C_0\ne 0$.
 Section~\ref{sec:cup} reveals the morphology of stable or metastable membrane patches
and proposes a $\bar{\kappa}$ estimation method using  these  membrane shapes.
Section~\ref{sec:tran} presents the transient shape of opening and closing membrane patches
and discusses  the difference in estimated $\bar{\kappa}$ values using the above two methods.
Section~\ref{sec:assem} describes the effects of $C_0$ on the membrane self-assembly dynamics.
Finally, Sec.~\ref{sec:sum} presents a summary of this study.

\section{Simulation model and method}~\label{sec:method}

The details of the meshless membrane model are described in Ref.~\onlinecite{shib11};
therefore, it is only briefly described here.
A fluid membrane is represented by a self-assembled one-layer sheet of $N$ particles.
The position and orientational vectors of the $i$-th particle are ${\bf r}_{i}$ and ${\bf u}_i$, respectively.
The membrane particles interact with each other via a potential $U=U_{\rm {rep}}+U_{\rm {att}}+U_{\rm {bend}}+U_{\rm {tilt}}$.
The potential $U_{\rm {rep}}$ is an excluded volume interaction with a diameter $\sigma$ for all pairs of particles.
The solvent is implicitly accounted for by an effective attractive potential 
\begin{eqnarray} \label{eq:U_att}
\frac{U_{\rm {att}}}{k_{\rm B}T} =  \frac{\varepsilon}{4}\sum_{i} \ln[1+\exp\{-4(p_i-p^*)\}]- b,
\end{eqnarray} 
with  $p_i= \sum_{j \ne i} f_{\rm {cut}}(r_{i,j})$, $b= 0.25\ln\{1+\exp(4p^*)\}$,
and $k_{\rm B}T$ is the thermal energy,
where $f_{\rm {cut}}(r)$ is a $C^{\infty}$ cutoff function
 and $r_{i,j}=|{\bf r}_{i,j}|$ with ${\bf r}_{i,j}={\bf r}_{i}-{\bf r}_j$:
\begin{equation} \label{eq:cutoff}
f_{\rm {cut}}(r)=\left\{ 
\begin{array}{ll}
\exp\{a(1+\frac{1}{(r/r_{\rm {cut}})^n -1})\}
& (r < r_{\rm {cut}}) \\
0  & (r \ge r_{\rm {cut}}) 
\end{array}
\right.
\end{equation}
Here, $n=12$, $a=\ln(2) \{(r_{\rm {cut}}/r_{\rm {att}})^n-1\}$,
$r_{\rm {att}}= 1.8\sigma$  $(f_{\rm {cut}}(r_{\rm {att}})=0.5)$, 
and the cutoff radius $r_{\rm {cut}}=2.1\sigma$.
The density $p^*=6$ in $U_{\rm {att}}(p_i)$ is the characteristic density.
For low density $p_i < p^*-1$,
it is a pairwise attractive potential, whereas the attraction is
smoothly truncated at $p_i \gtrsim p^*$.
This truncation allows the formation of a fluid membrane over a wide range of parameters.
The bending and tilt potentials
are given by 
\begin{eqnarray}
\frac{U_{\rm {bend}}}{k_{\rm B}T} &=& \frac{k_{\rm {bend}}}{2} \sum_{i<j} ({\bf u}_{i} - {\bf u}_{j} - C_{\rm {bd}} \hat{\bf r}_{i,j} )^2 w_{\rm {cv}}(r_{i,j}), \\
\frac{U_{\rm {tilt}}}{k_{\rm B}T} &=& \frac{k_{\rm{tilt}}}{2} \sum_{i<j} [ ( {\bf u}_{i}\cdot \hat{\bf r}_{i,j})^2
 + ({\bf u}_{j}\cdot \hat{\bf r}_{i,j})^2  ] w_{\rm {cv}}(r_{i,j}), 
\end{eqnarray}
where
 $\hat{\bf r}_{i,j}={\bf r}_{i,j}/r_{i,j}$ and $w_{\rm {cv}}(r_{i,j})$ is a weight function.

The membrane elastic properties are given in Ref.~\onlinecite{shib11} for a wide range of parameter sets.
The bending rigidity $\kappa$ is linearly dependent on $k_{\rm {bend}}$ and $k_{\rm{tilt}}$.
Here, we fix the ratio as $k_{\rm {bend}}=k_{\rm{tilt}}=k$.
The spontaneous curvature, $C_0$, of the membrane is 
given by $C_0\sigma= C_{\rm {bd}}/2$.
The line tension, $\gamma$, of the membrane edge is linearly dependent on $\varepsilon$.
When the thermal fluctuation is not very large ($\kappa/k_{\rm B}T \gtrsim 20$), 
$\kappa$ and $\gamma$ are independent of  $\varepsilon$ and
$k$, respectively, so that these two quantities are controlled individually.
The values of $\kappa/k_{\rm B}T= 16 \pm 1$ ($17 \pm 2$), $34 \pm 1$ ($35 \pm 1$), 
$52 \pm 3$ ($53 \pm 3$), and $69 \pm 3$ ($71 \pm 3$) are obtained
for  $k=10$, $20$, $30$, and $40$ at $\varepsilon=4$ ($5$), respectively, 
from the fluctuation spectrum of flat membranes.
Moreover, $\gamma\sigma/k_{\rm B}T= 3.89 \pm 0.04$ and $5.08 \pm 0.04$ 
are obtained for $\varepsilon=4$ and $5$ 
 at $k=30$ from the membrane strips 
and the estimated values at $k=20$ and $40$ are within their statistical errors.
These values are in the same range as those of lipid membranes.

Molecular dynamics with a Langevin thermostat is employed~\cite{shib11,nogu11}.
To calculate the closing probability, $500$ independent runs are performed for each pre-curved membrane.
The pre-curved membrane is prepared by adding the constraint potential 
on a sphere $U_{\rm sp}/k_{\rm B}T=(k_{\rm sp}/2)\sum_i (r_i-r_{\rm int})^2$, where $r_i$ is the distance of the $i$-th particle from the origin ($r_i=|{\bf r}_i|$).
If not specified, $k_{\rm sp}=0.2$ is used.

In the self-assembly simulation, 
the number $N$ of membrane particles is varied; {\it i.e.,} $N=5000$, $10~000$, $20~000$, and $40~000$
in the cubic simulation box with the side length $L=128\sigma$ with the periodic boundary conditions:
the density  $\phi=N/L^3=0.0024$, $0.0048$, $0.0095$, and $0.019$, respectively.
The numerical errors of the mean cluster size are calculated using the standard errors of 20 independent runs.
In the following, the results are displayed with the diameter of membrane particles, $\sigma$, as the length unit
 and $\tau= \sigma^2/D_{\rm mono}$ as the time unit,
where $D_{\rm mono}=k_{\rm B}T/\zeta$ is the diffusion coefficient of isolated membrane particles,
and $\zeta$ is the friction constant of the Langevin thermostat.

\begin{figure}
\includegraphics{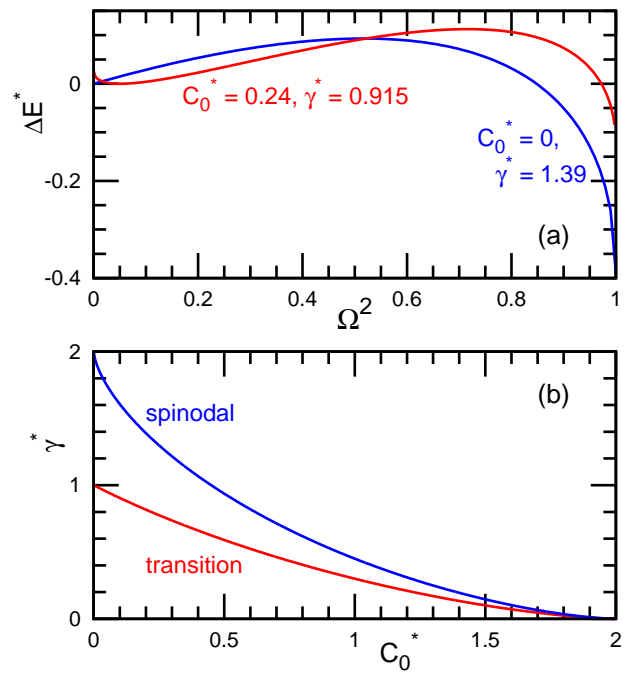}
\caption{
Theoretical prediction of spherical-cap-shaped membranes.
(a) Energy profile for membrane closure given by Eq.~(\ref{eq:en})
for $\{C_0^*, \gamma^*\}=\{0, 1.39\}$ and $\{0.24, 0.915\}$.
(b) Phase diagram. The vesicle and cup-shaped membrane patches 
exist in thermal equilibrium above and below the transition curve, respectively.
The energy barrier between the two states disappears above the spinodal curve.
}
\label{fig:theory}
\end{figure}

\section{Theory}\label{sec:theory}

The energy of a membrane patch is expressed by
\begin{equation}
E = \int \Big[ \frac{\kappa}{2}(C_1+C_2 -  C_0)^2 + \bar{\kappa}C_1C_2 \Big] dA + \oint \gamma ds,
\end{equation}
 where $C_1$ and $C_2$ are the two principal curvatures of the membrane~\cite{canh70,helf73}.
The first and second integrals are calculated over the membrane area and along the membrane boundary, respectively.
According to previous studies~\cite{from83,hu12}, it is approximated
 that the membrane geometry is a spherical cap.
In cylindrical coordinates, the spherical cap is represented by $\{\rho, z\}= \{r\sin(\theta), r\cos(\theta)\}$ 
where $\rho^2=x^2+y^2$, angle $\theta=[\theta_{\rm {ed}},\pi]$, and the membrane area $A= 2\pi r^2 (1+\cos(\theta_{\rm ed}))$.
The length is normalized by the radius of the vesicle, $R_{\rm ves}= \sqrt{A/4\pi}$ and
the membrane curvature is normalized as $\Omega= R_{\rm ves}/r= \sqrt{(1+\cos(\theta_{\rm ed}))/2}$.
Then, the energy is given by
\begin{equation}
E^*(\Omega^2,C_0^*,\gamma^*) =  \Omega^2 - C_0^* \Omega + \gamma^*\sqrt{1-\Omega^2}, \label{eq:en}
\end{equation}
where $E^*= E/4\pi (2\kappa+\bar{\kappa})$,
$C_0^* = 2\kappa C_0 R_{\rm ves}/(2\kappa+\bar{\kappa})$, and $\gamma^*= \gamma R_{\rm ves}/(2\kappa+\bar{\kappa})$.
The constant term, $\kappa{C_0}^2R_{\rm ves}^2/2(2\kappa+\bar{\kappa})$, is neglected.
For a symmetric membrane ($C_0^*=0$), Eq.~(\ref{eq:en}) coincides with the equations in Refs.~\onlinecite{from83,hu12}.
For $\gamma^*<1$ and $\gamma^*>1$, a flat disk ($\Omega=0$) and a vesicle ($\Omega=1$) are the global energy minima,
respectively.
Thus, the disk-to-vesicle transition occurs at $\gamma^*=1$.
For $0<\gamma^*<2$, an energy barrier exists between the disk and vesicle so that the two states can coexist.
As the membrane patches grow in the self-assembly dynamics,
vesicles are formed via nucleation at $1<\gamma^* \lesssim 2$
and via spinodal decomposition at $\gamma^* \gtrsim2$ [$R_{\rm ves} \gtrsim 2(2\kappa+\bar{\kappa})/\gamma$].

Figure~\ref{fig:theory}(a) shows the energy profiles for $C_0^*=0$ and $C_0^*=0.24$
where the energy minimum, $E_{\rm cup}^*$, of the cup-shaped (or flat at $C_0^*=0$) membrane patch is set to the origin as
$\Delta E^*= E^*- E_{\rm cup}^*$.
As explained above, two minima exist at $\Omega=0$ and $1$ at $C_0^*=0$.
At a finite value of $C_0^*$,
 the minimum of $\Omega=0$ shifts to the right so that the membrane patch becomes a cup shape.
As $C_0^*$ increases, the transition and spinodal points decrease, as shown in Fig.~\ref{fig:theory}(b).

We obtained the curvature of the energy-minimum cup-shape using Taylor expansion as
\begin{equation}
\Omega_{\rm cup}= \frac{C_0^*}{2-\gamma^*} + \frac{\gamma^* {C_0^*}^3}{2(2-\gamma^*)^4} + \frac{3\gamma^*(2+\gamma^*) {C_0^*}^5}{8(2-\gamma^*)^7} + O({C_0^*}^7).  \label{eq:cup}
\end{equation}
The deviations of this approximation from the exact values of $\Omega_{\rm cup}$ are less than 5\% 
in our simulation conditions: 0.03\% and 3\% for $C_0^*=0.2$ and $0.4$ at $\gamma^*=1$, respectively.

\begin{figure}
\includegraphics{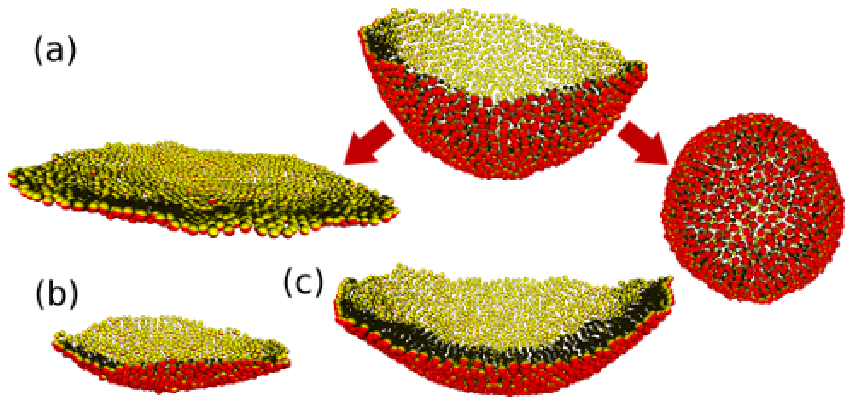}
\includegraphics{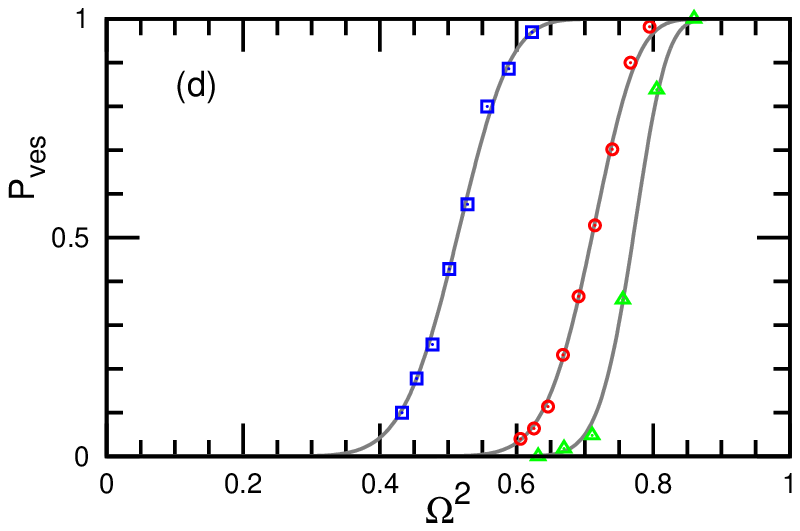}
\caption{
Shape changes of pre-curved membrane patches.
(a)--(c) Membrane snapshots. 
The orientation vector of a membrane particle lies along the direction from the 
light (yellow) to dark gray (red) hemispheres.
The middle snapshot in (a) shows an initial membrane conformation.
The left snapshot in (a) and those in (b) and (c)
show the disk- and cup-shaped membrane patches in their final conformations, respectively.
(d) Closing probability, $P_{\rm {ves}}$.
The snapshots in (a) and squares in (d) represent
the data for $C_0=0$, $N=1600$, $k=20$, and $\varepsilon=4$.
The snapshot in (b)  and circles in (d) represent the data for $C_0\sigma=0.015$, $N=800$, $k=20$, and $\varepsilon=4$.
The snapshot in (c)  and triangles in (d) represent the data for $C_0\sigma=0.025$, $N=1600$, $k=40$, and $\varepsilon=4$.
Three solid lines are obtained through fitting to Eq.~(\ref{eq:close}).
}
\label{fig:close}
\end{figure}

\begin{figure}
\includegraphics{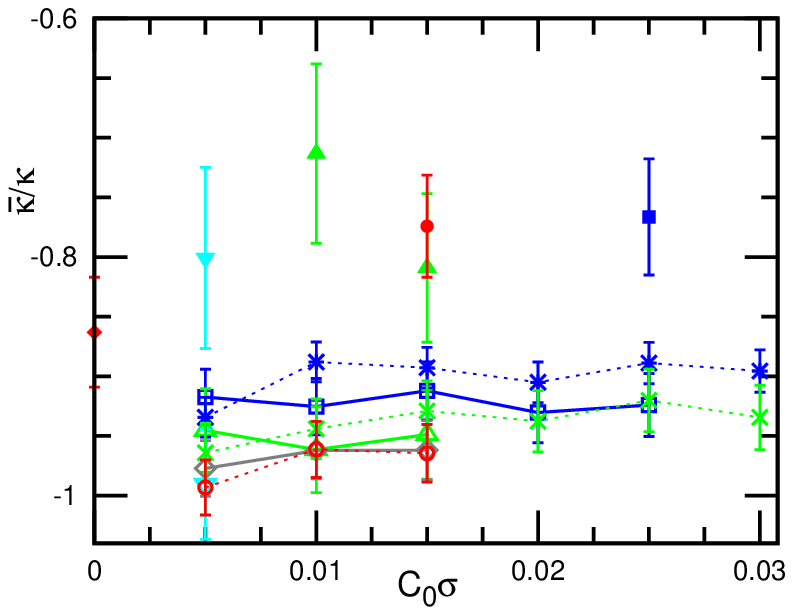}
\caption{
Estimation of the Gaussian modulus $\bar{\kappa}$.
The closed symbols represent $\bar{\kappa}/\kappa$ estimated from the closing probability
for ($\bullet$)  $N=800$ and ($\blacklozenge$, $\blacktriangle$, $\blacktriangledown$, $\blacksquare$) $N=1600$.
The open symbols with dashed and solid lines represent $\bar{\kappa}/\kappa$ 
estimated from the shapes of cup-shaped membrane patches
 at $N=800$ and $1600$, respectively.
$\bullet$, $\circ$, $\blacklozenge$: $k=20$ and $\varepsilon=4$.
$\blacktriangle$, $\triangle$, $\times$: $k=30$ and $\varepsilon=4$.
$\blacksquare$, $\square$, $\divideontimes$:  $k=40$ and $\varepsilon=4$.
$\blacktriangledown$, $\triangledown$: $k=30$ and $\varepsilon=5$.
}
\label{fig:kgr}
\end{figure}

\begin{figure}
\includegraphics{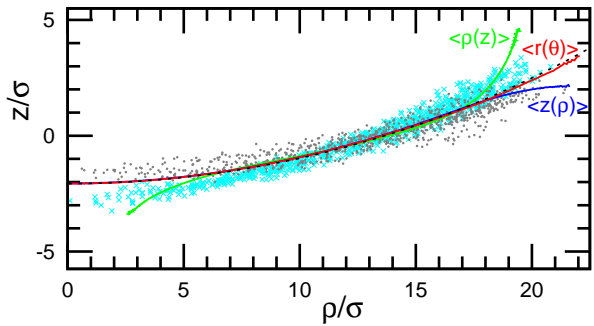}
\caption{
Conformation of the metastable cup-shaped membrane patch in the cylindrical coordinate $(\rho, z)$ 
at the parameter set of (b) in Fig.~\ref{fig:close}.
Two examples of instantaneous conformation 
(membrane particle positions) are shown in symbols $\bullet$ and $\times$.
Three solid lines represent the averages 
$\langle \rho(z) \rangle$,  $\langle r(\theta) \rangle$, and  $\langle z(\rho) \rangle$. 
The dashed line represents the fit to a circle.
}
\label{fig:cup}
\end{figure}

\section{Closing Probability} \label{sec:close}

When an energy barrier exists between the cup-shaped membrane and vesicle [below the spinodal curve in Fig.~\ref{fig:theory}(b)],
the  pre-curved membrane with a spherical-cap shape of $\Omega_{\rm cup}<\Omega<1$ transforms into either state.
Hu {\it et al.}~\cite{hu12} derived this  probability for a zero-spontaneous-curvature membrane
to estimate the Gaussian modulus, $\bar{\kappa}$, of membranes.
First,  $\bar{\kappa}$ was calculated for a solvent-free molecular model~\cite{hu12} and 
later for the MARTINI model~\cite{hu13} and a dissipative-particle-dynamics model\cite{naka15}.
However, it has not been applied to a membrane with a finite spontaneous curvature.
Here, we extended it for $C_0 \neq 0$ in a straightforward manner:
\begin{equation}
P_{\rm ves}(\Omega^2)= \frac{\int_{\Omega_{\rm cup}^2}^{\Omega^2} \exp(\frac{\Delta E(u, C_0^*,\gamma^*)}{D^*}) du}{\int_{\Omega_{\rm cup}^2}^1  \exp(\frac{\Delta E(u, C_0^*,\gamma^*)}{D^*})du}, \label{eq:close}
\end{equation}
where the normalized diffusion constant $D^*=D/(2\kappa+\bar{\kappa})$.
Although   $P_{\rm ves}(\Omega^2)$ can be analytically calculated for $C_0 = 0$,~\cite{hu12}
it must be numerically calculated for $C_0 \neq 0$.

First, we calculated the closing probability, $P_{\rm {ves}}$, of a symmetric membrane ($C_0=0$) at $N=1600$, $k=20$, and $\varepsilon=4$
that has $\kappa/k_{\rm B}T= 34 \pm 1$, $\gamma\sigma/k_{\rm B}T= 3.89 \pm 0.04$, and $R_{\rm ves}/\sigma=13.81 \pm 0.01$, 
as shown in Fig.~\ref{fig:close}(a) and by squares in Fig.~\ref{fig:close}(d).
It fits well to Eq.~(\ref{eq:close}) with $\gamma^*=1.39$ and $D^*=0.004$ as the solid line overlaps with the squares in Fig.~\ref{fig:close}(d).
At this value of $\gamma^*$, the vesicle ($\Omega=1$) has lower energy than the disk-shaped patch ($\Omega=0$)
as $E_{\rm disk}^*-E_{\rm ves}^* = \gamma^*-1=0.39$ and the energy barrier of $E_{\rm max}^*-E_{\rm disk}^* =0.093$ exists as shown in  Fig.~\ref{fig:theory}(a). 
The Gaussian modulus is obtained as $\bar{\kappa}/\kappa= -0.86 \pm 0.05$,
 where the estimation errors are calculated from the errors of $\kappa$ and $\gamma$.
This value is in the range of the  Gaussian modulus reported in the previous experiments and molecular simulations $\bar{\kappa}/\kappa \simeq -1$. ~\cite{hu12,hu13,naka15}

Next, we calculated the  $P_{\rm {ves}}$ value of asymmetric membranes ($C_0>0$) for five parameter 
sets, as shown by the closed symbols in Fig.~\ref{fig:kgr}.
For all sets, $P_{\rm {ves}}$ fits well to Eq.~(\ref{eq:close}). 
In Fig.~\ref{fig:close}(d), only two of them are shown for clarity.
Equation~(\ref{eq:close}) has two fitting parameters, $\bar{\kappa}/\kappa$ and $D^*$, which predominantly
 determine the position and sharpness of the $P_{\rm {ves}}$ increase, respectively.
Here, the value of $D$ obtained at $C_0=0$ (squares in Fig.~\ref{fig:close}(d))
 is used so that the dimensionless value is $D^*/D_0^*=\kappa_0/\kappa$ where $D_0^*=0.004$ and $\kappa_0/k_{\rm B}T=34$.
Thus, the diffusion constant is independent of the potential parameters in the presented simulations.
This is reasonable because membrane closing and opening are normal membrane motions that are much less dependent on the potential interactions than the tangential motion in the meshless membrane model.
We fit $\bar{\kappa}/\kappa$  through  minimization of 
$\sum_i (P_{\rm {ves}}^{\rm fit}(\Omega^2_i)-P_{\rm {ves}}^{\rm sim}(\Omega^2_i))^2$.
In a general case, $D^*$ can be obtained by fitting.

Thus, the closing probability, $P_{\rm {ves}}$, is well expressed by Eq.~(\ref{eq:close}).
The following sections 
describe (meta)stable and transient membrane shapes
and discuss the accuracy of $\bar{\kappa}$ estimation.

\section{Cup-Shaped Membrane Patch} \label{sec:cup}

At $C_0>0$, a cup-shaped membrane patch is formed as a stable or metastable state 
below the transition curve and between the transition and spinodal curves
in Fig.~\ref{fig:theory}(b), respectively.
We clarified that these shapes are well described by spherical-cap shapes such that
the Gaussian curvature can be estimated from them.

Figure~\ref{fig:cup} shows the membrane conformation of the metastable cup-shaped membrane patch
at the same condition as the data given in Fig.~\ref{fig:close}(b) and the circles in Fig.~\ref{fig:close}(d).
The center of the mass of the membrane patch is taken to the origin
and the $z$ axis is taken along the eigenvector of the smallest eigenvalue of the gyration tensor.
Then, the membrane positions are averaged in three different ways in the cylindrical coordinate $(\rho, z)$
to obtain the equilibrium membrane shape.
The first two averages are obtained by taking the average of each $z$ or $\rho$ coordinate as
$\langle \rho(z) \rangle$ and $\langle z(\rho) \rangle$, respectively.
A long-time average of the particle height $z_i$ in a narrow bin of $\rho- \Delta\rho/2 \leq \rho_i < \rho+ \Delta\rho/2$
is calculated as $\langle z(\rho) \rangle$. Similarly, $\langle \rho(z) \rangle$ is calculated 
for $z-\Delta z/2 \leq z_i < z+ \Delta z/2$. Here, $\Delta \rho= 0.025\sigma$ and $\Delta z= 0.05\sigma$ are used.
Since the membrane conformation substantially fluctuates,
the curves of $\langle \rho(z) \rangle$ and $\langle z(\rho) \rangle$ bends in the upper and lower directions, 
respectively, at the ends of the curves in Fig.~\ref{fig:cup}.
Compared to these two averages, $\langle z(\rho) \rangle$ gives a better description for $\rho\simeq \sigma$.
Therefore, the curvature radius, $r_{\rm cup}$, of the membrane can be calculated by a least-squares fit to a circle using minimization 
of $\Lambda= (1/2)\sum [\rho^2+(z(\rho)-z_{\rm G})^2-r_{\rm cup}^2]^2$ for $\rho\lesssim 18\sigma$ as
\begin{eqnarray}
r_{\rm cup}^2 &=& \langle \rho^2 \rangle + \langle z^2 \rangle - 2\langle z \rangle z_{\rm G}   + z_{\rm G}^2, \\
z_{\rm G} &=& \frac{ \langle r^2 z\rangle - \langle r^2 \rangle \langle z \rangle }{2(\langle z^2 \rangle - \langle z\rangle^2)},
\end{eqnarray}
where $r^2=\rho^2+z^2$.

Then, the third average taken for the radius $r(\theta)$ 
of the fitted circle $r=\sqrt{\rho_i^2+(z_i-z_{\rm G})^2}$ was calculated,
where $(\rho_i,z_i)=(r\sin(\theta), -r\cos(\theta)+z_{\rm G})$.
The average $\langle r(\theta) \rangle$ gives a better estimate up to larger $\rho$.
The radius, $r_{\rm cup}$, can be also calculated from the fit to  $\langle r(\theta) \rangle$.
The latter fit was used in this study, although the difference between the two fits is less than their statistical errors.
The fitted curve (dashed line) coincides well with $\langle r(\theta) \rangle$, as shown in Fig.~\ref{fig:cup}.
Thus, the membrane shape is well expressed by a spherical cap.
Using the obtained curvature  $1/r_{\rm cup}$ and  Eq.~(\ref{eq:cup}) with $\Omega= R_{\rm ves}/r_{\rm cup}$, 
$\bar{\kappa}/\kappa$ is estimated as shown in Fig.~\ref{fig:kgr}.
This method can be applied to a wider range of membrane sizes than the former closing-probability method.
The former method requires a sufficiently high transition barrier between the cup-shaped (or disk-shaped) 
membrane patches and vesicle; therefore, 
 the membrane size must be carefully selected.
In contrast, the latter method does not require a barrier and can be applied to smaller size membranes.

The obtained values of $\bar{\kappa}/\kappa$ are almost constant ($-0.9\pm 0.1$) for all parameter sets 
but are lower than the results of the closing probability.
This inconsistency is larger than numerical errors.
When the value of $\bar{\kappa}/\kappa$ obtained from $r_{\rm cup}$ is used,
$P_{\rm {ves}}$ is shifted to the left: $\Delta\Omega^2=0.17$ and $0.18$ 
for the data indicated by the circles and triangles in Fig.~\ref{fig:close}(d).
The next section describes transient membrane shapes
and discusses the reason for this difference.

\begin{figure}
\includegraphics{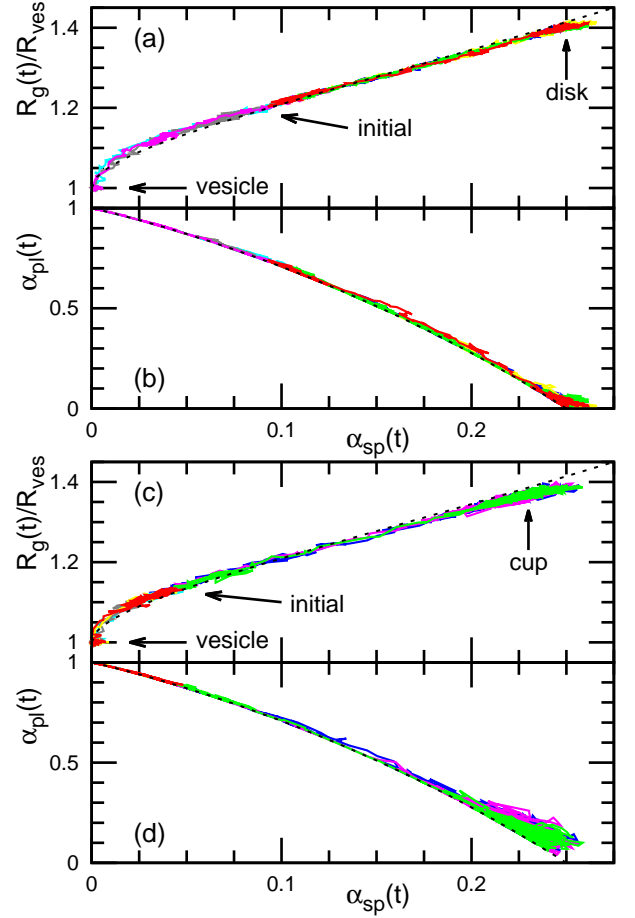}
\caption{
Trajectories of asphericity, $\alpha_{\rm sp}$, and aplanarity, $\alpha_{\rm pl}$, from pre-curved membranes.
The upper two panels (a),(b) and lower two panels (c),(d) show the data
for the parameter sets of (a) and (b) in Fig.~\ref{fig:close}, respectively.
Eight trajectories are shown for each parameter set.
The dashed lines represent the spherical-cap shape given by Eqs.~(\ref{eq:cap1}) and (\ref{eq:cap2}).
}
\label{fig:traj}
\end{figure}

\section{Transient membrane shapes} \label{sec:tran}

To quantify the transient shapes of the closing and opening membranes in the closing-probability calculations,
we calculate three shape parameters based on
the gyration tensor $a_{\alpha\beta}= (1/N)\sum_i (\alpha_{i}-\alpha_{\rm G})(\beta_{i}-\beta_{\rm G})$
where $\alpha,\beta\in \{x,y,z\}$:
\begin{eqnarray}
R_{\rm g}^2 &=& T_{\rm {gy}} = \lambda_1+\lambda_2+\lambda_3, \\ \nonumber
\alpha_{\rm {sp}} &=& 1- \frac{3M_{\rm {gy}}}{T_{\rm {gy}}^2} = \frac{({\lambda_1}-{\lambda_2})^2 + 
  ({\lambda_2}-{\lambda_3})^2+({\lambda_3}-{\lambda_1})^2}{2(\lambda_1+\lambda_2+\lambda_3)^2}, \\ \nonumber
\alpha_{\rm {pl}} &=& \frac{9D_{\rm {gy}}}{T_{\rm {gy}}M_{\rm {gy}}}  = \frac{9\lambda_1\lambda_2\lambda_3} {(\lambda_1+\lambda_2+\lambda_3)
    (\lambda_1\lambda_2+\lambda_2\lambda_3+\lambda_3\lambda_1)},
\end{eqnarray}
where  $T_{\rm {gy}}$, $D_{\rm {gy}}$, and $M_{\rm {gy}}$ are  three invariants of the gyration tensor:
trace, determinant, and the sum of its three minors ($M_{\rm {w}}= a_{xx}a_{yy}+a_{yy}a_{zz}+a_{zz}a_{xx}
-a_{xy}^2-a_{yz}^2-a_{zx}^2$) for the gyration tensor, respectively.
These quantities can be calculated directly from three invariants
without calculating the eigenvalues, ${\lambda_1} \leq {\lambda_2} \leq {\lambda_3}$.
Spherical, disk, and rod shapes are well distinguished by $\alpha_{\rm {sp}}$ and $\alpha_{\rm {pl}}$. 
The asphericity, $\alpha_{\rm {sp}}$, indicates the shape deformation from a spherical 
shape to a thin-rod shape:~\cite{rudn86}  $\alpha_{\rm {sp}}=0$, $0.25$, and $1$ 
for a sphere (${\lambda_1}={\lambda_2}={\lambda_3}$), thin disk (${\lambda_1}=0, {\lambda_2}={\lambda_3}$), 
and thin rod (${\lambda_1}={\lambda_2}=0$), respectively.
A discocyte (red-blood-cell) shape takes $\alpha_{\rm {sp}}\simeq 0.2$.~\cite{nogu05}
The aplanarity, $\alpha_{\rm {pl}}$, indicates the deviation from a plane:~\cite{nogu06}
 $\alpha_{\rm {pl}}=0$  and $1$ for a flat membrane (${\lambda_1}=0$) and sphere, respectively.
The transition between flat and spherical arrangements was quantified by $\alpha_{\rm {pl}}$ in Ref.~\onlinecite{nogu16b}.

For a spherical cap, these quantities are given by
\begin{eqnarray}
\label{eq:rgas1}
\frac{R_{\rm {g}}}{R_{\rm {ves}}} &=&
       \sqrt{\frac{3-\cos(\theta_{\rm {ed}})}{2}}, \\
\alpha_{\rm {sp}}  &=& 
\frac{(1-\cos(\theta_{\rm {ed}}))^2}{(3-\cos(\theta_{\rm {ed}}))^2},
\label{eq:rgas2} \\
\alpha_{\rm {pl}}  &=& 
 2+\cos(\theta_{\rm {ed}}) - \frac{4}{3-\cos(\theta_{\rm {ed}})},
\label{eq:rgas3}
\end{eqnarray} 
which imply that
\begin{eqnarray} \label{eq:cap1}
\alpha_{\rm {sp}} &=& \bigg[1-  \Big(\frac{R_{\rm {ves}}}{R_{\rm {g}}}\Big)^2 \bigg]^2, \\ \label{eq:cap2}
\alpha_{\rm {pl}} &=& \frac{(1-2\sqrt{\alpha_{\rm {sp}}})(1+\sqrt{\alpha_{\rm {sp}}})}{1-\sqrt{\alpha_{\rm {sp}}}}.
\end{eqnarray} 
Equation~(\ref{eq:cap1}) was derived in Ref.~\onlinecite{nogu06a} and
used to clarify that the membrane closes via an oblate shape at $\gamma^* \gg 1$ 
while it is via a spherical cap for $\gamma^* \sim 1$ at $C_0=0$.
Here, the transient membrane shapes are examined using Eq.~(\ref{eq:cap1}) and also Eq.~(\ref{eq:cap2}).

\begin{figure}
\includegraphics{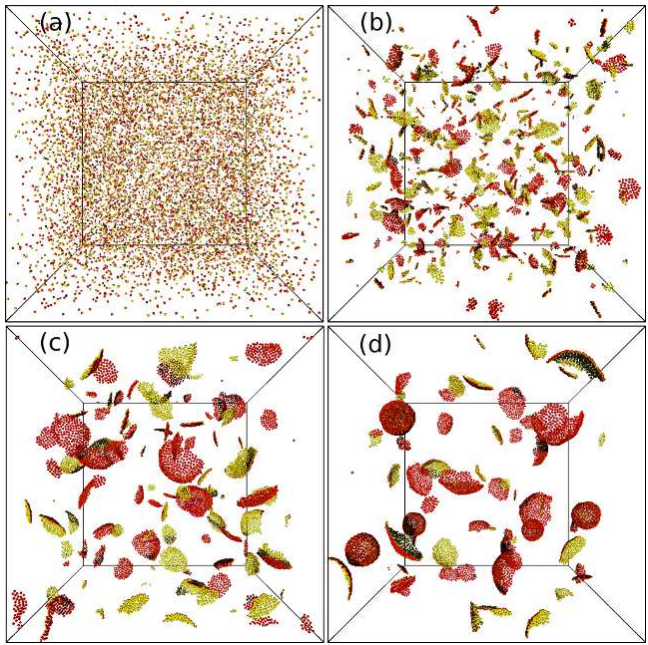}
\caption{
Sequential snapshots of membrane self-assembly for  $C_0\sigma=0.05$,  $\phi= 0.0048$, $k=20$, and $\varepsilon=4$
 at (a) $t/\tau=0$, (b)  $t/\tau=5000$, (b)   $t/\tau=25~000$, and (d)   $t/\tau=50~000$.
}
\label{fig:snapc1}
\end{figure}

\begin{figure}
\includegraphics[width=8cm]{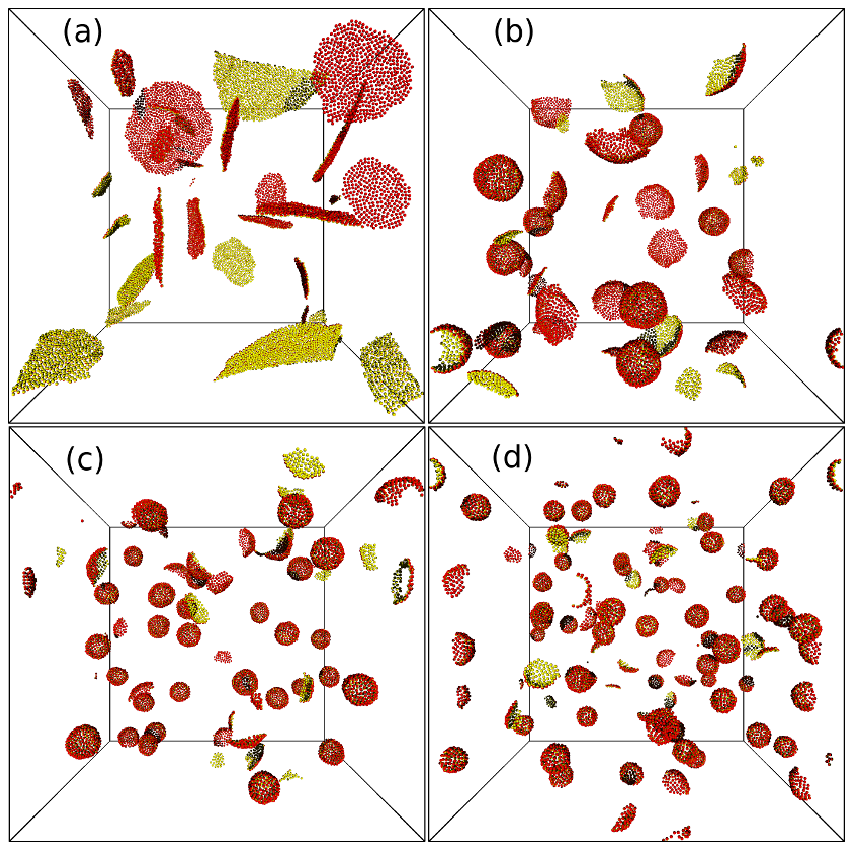}
\includegraphics{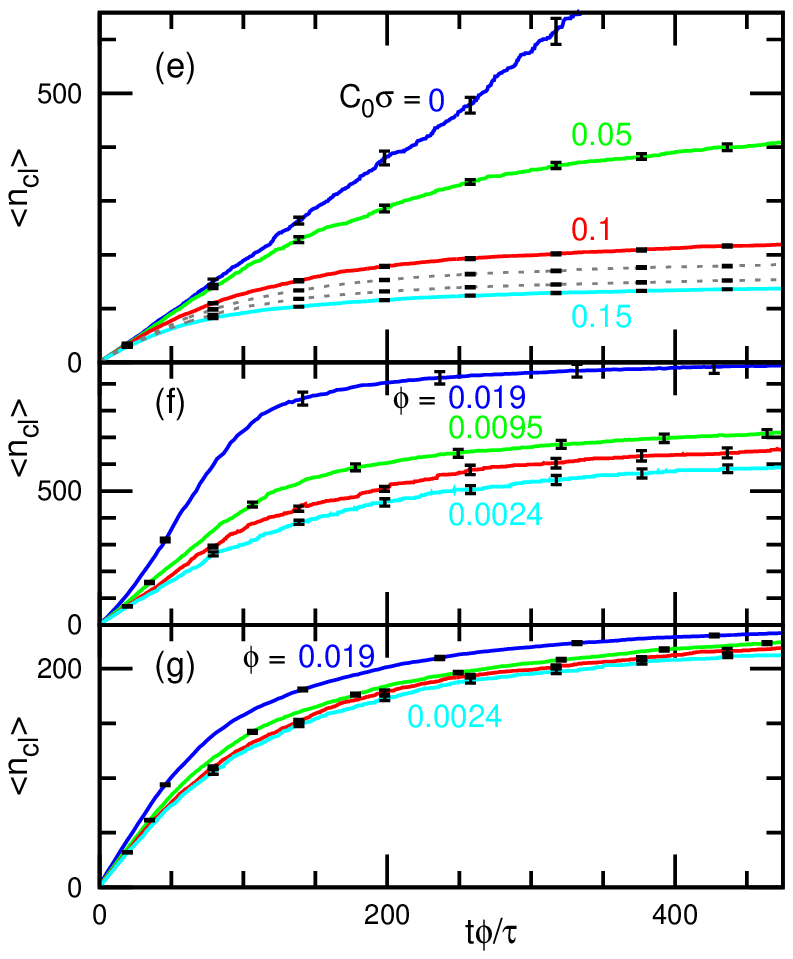}
\caption{
Membrane self-assembly.  
(a)--(d) Snapshots  at $t/\tau=100~000$ for $\phi= 0.0048$, $k=20$, and $\varepsilon=4$.
(a) $C_0=0$. (b) $C_0\sigma=0.05$. (c) $C_0\sigma=0.1$.  (d) $C_0\sigma=0.15$. 
(e) Temporal evolution of mean cluster size $\langle n_{\rm cl} \rangle$ at  $\phi=0.0048$ and $k=20$.
The solid lines in (e) represent the data for $C_0\sigma=0$, $0.05$, $0.1$, and $0.15$ 
at $\varepsilon=4$.
The upper and lower dashed lines in (e) represent the data for $\varepsilon=5$ and $6$, respectively, at $C_0\sigma=0.1$.
(f), (g) Temporal evolution of $\langle n_{\rm cl} \rangle$ for $\phi=0.0024$, $0.0048$, $0.0095$, and $0.019$
(f) at $C_0=0$, $k=10$, and $\varepsilon=6$ and (g) at $C_0\sigma=0.1$, $k=20$, and $\varepsilon=4$.
The error bars are displayed at several data points.
}
\label{fig:cnm}
\end{figure}

Figure~\ref{fig:traj} shows the trajectories for $C_0\sigma=0$ and $0.015$, which
 change shapes along the dashed lines of Eqs.~(\ref{eq:cap1}) and (\ref{eq:cap2}); thus,
 the transient shapes are well described by the spherical cap.
However, fluctuations around the metastable-cup state are recognizable at $C_0\sigma=0.015$ [see Figs.~\ref{fig:traj}(c) and (d)].
Since the disk and cup-shaped membrane patches have an open edge, they can fluctuate more than the vesicle and have more conformational entropy.
This entropy increases the stability of the opened states and shifts the closing probability $P(\Omega^2)$ to the right
so that it induces the upper shifts of $\bar{\kappa}/\kappa$ in Fig.~\ref{fig:kgr}.
We also check that the closing probability and trajectories are not modified by the initial constraint  
by examining a stronger constraint with $k_{\rm sp}=1$.
Thus, it is concluded that the amplitude of $\bar{\kappa}/\kappa$ 
estimated from the closing probability is a slight underestimate.

\section{Self-assembly} \label{sec:assem}

Next, we investigate how the cup-to-vesicle transition occurs during the membrane self-assembly.
Figures~\ref{fig:snapc1} and \ref{fig:cnm} show the  membrane self-assembly from random gas states.
First, the  particles assemble into small cup-shaped membrane patches, and then
the membrane patches fuse into larger patches;
sufficiently large patches close into vesicles.
The vesicles are not fused since  an energy barrier is required for this topological change.
Thus, the final vesicle size is determined by the membrane closure.
As expected, smaller vesicles are formed at higher $C_0$ [see Figs.~\ref{fig:cnm}(a)--(d)].
To quantify these phenomena, the mean cluster size 
$n_{\rm {cl}}= (1/N)\sum_{i_{\rm cl}=1}^{\infty} i_{\rm cl}^2 n^{\rm cl}_i$ was calculated,
where $n^{\rm cl}_i$ is the number of clusters of size $i_{\rm cl}$
and $N=\sum_{i_{\rm {cl}}=1}^{\infty} i_{\rm {cl}} n^{{\rm {cl}}}_i$.   
It is considered that particles belong to a
cluster when their distance is less than $r_{\rm {att}}$.
As reported in Ref.~\onlinecite{nogu06a} for $C_0=0$,
$\langle n_{\rm {cl}}\rangle$ linearly increases with time for disk-shaped patches
and saturates via the vesicle formation [see Figs.~\ref{fig:cnm}(e)--(g)].
These dynamics do not qualitatively change for $C_0\neq0$.

Using $\bar{\kappa}/\kappa \simeq -0.9$, 
the transition and spinodal sizes $\{N_{\rm tra}, N_{\rm spi}\}\simeq \{770, 3100\}$, 
$\{260,440\}$, $\{140,210\}$, and $\{90, 130\}$ were obtained
for  $C_0\sigma=0$, $0.05$, $0.1$, and $0.15$, respectively, 
at $k=20$ and $\varepsilon=4$ [solid lines in Fig.~\ref{fig:cnm}(e) and Fig.~\ref{fig:cnm}(g)];
and $\{N_{\rm tra}, N_{\rm spi}\}\simeq \{160, 650\}$ at $C_0=0$, $k=10$, and $\varepsilon=6$
[Fig.~\ref{fig:cnm}(f)].
These threshold sizes are consistent with the simulation results; 
the vesicle formation occurs at  $i_{\rm cl}\gtrsim N_{\rm spi}$.
As $C_0$ increases, the vesicles form faster:
the mean time of the first vesicle formation is $210~000\tau$, $21~000\tau$, $10~000\tau$, and $6600\tau$  
at $C_0\sigma=0$, $0.05$, $0.1$, and $0.15$, respectively, 
for the solid lines in Fig.~\ref{fig:cnm}(e).
As the edge tension ($\varepsilon$) increases, 
the vesicles formation starts earlier and 
smaller vesicles are obtained [see dashed lines in Fig.~\ref{fig:cnm}(e)].

The effects of the density, $\phi$, are shown in Figs.~\ref{fig:cnm}(f) and (g). 
Since the collision frequency is proportional to $\phi$  in a dilute solution,
 $\langle n_{\rm {cl}}\rangle$ initially increases as $t\phi$.
Interestingly, the vesicle size increases with increasing  $\phi$.
Since the membrane closure takes time,
the fusion of large patches with others more frequently occurs at higher $\phi$ before membrane closure.
This closing time is $\sim 1000\tau$ and  $\sim 100\tau$ for the case shown in Figs.~\ref{fig:cnm}(f) and (g), 
respectively; thus, larger $\phi$ dependence is seen in Fig.~\ref{fig:cnm}(f).

\section{Summary} \label{sec:sum}

We have studied the closing transition of membranes and their self-assembly at finite spontaneous curvatures.
The membrane shapes at the transition are well expressed by a spherical cap similar to zero-spontaneous-curvature membranes.
This validates the spherical-cap approximation in the theory.
We estimated the  Gaussian modulus from the closing probability and the curvature of (meta)stable cup-shaped membrane patches.
The former is a straightforward extension of the method proposed by Hu {\it et al.} for $C_0=0$.
The latter is newly proposed in this study and can be calculated for smaller size membranes.
The former gives slightly smaller negative values of the Gaussian modulus than the latter.
This underestimate is due to the higher conformational entropy of opened membranes with longer edges.
Thus, the latter method gives a better estimation,
 although it is available only for the non-zero spontaneous-curvature membranes.

During the self-assembly, vesicles are formed by the closure of the membrane patches, which
 occurs above the spinodal curves.
Therefore, the vesicle size is predominantly determined energetically by the spontaneous curvature,
bending rigidity, and edge line tension.
However, it also kinetically depends on the density, 
because the membrane patches can fuse with others before the closure is completed at high density.

We estimated the Gaussian modulus of the meshless membrane model for a wide range of parameters.
The result is $\bar{\kappa}/\kappa = -0.9 \pm 0.1$, which
is independent of $C_0$ and the parameter choices in the range of this study.
The values of the other membrane properties  have been thoroughly investigated by previous studies.
Since the membrane properties are well determined and tunable,
 the presented model is well suited for studying the membrane phenomena accompanying large shape deformation and topological changes.

Theoretically, we do not consider
the curvature energy of the membrane edges~\cite{morr19} and thermal fluctuation of membrane shapes.
Thus, the accuracy of the Gaussian modulus may be improved by taking them into account, particularly in the closing-probability method.
For much larger membrane patches than the thresholds $\gamma^* \gg 1$, transient shapes are oblate and multiple vesicles can be formed
depending on initial conformations~\cite{nogu09,nogu06a}.
Vesicle opening is observed  during lysis in a detergent solution~\cite{nomu01,nogu12a} and by the phase separation in charged lipid membranes~\cite{hime15}. Moreover, molecules of lower membrane edge tension are accumulated on the membrane edge. 
Extensions of this theory to non-spherical shapes and 
multi-component membranes represent an interesting problem for further research.

\begin{acknowledgments}
This work was supported by JSPS KAKENHI Grant Number JP17K05607.
\end{acknowledgments}

\end{document}